\documentclass{article}
\input{tcilatex}

\begin{document}

\textbf{IMAGINARY FIELDS}

\bigskip

E. A. Novikov

Institute for Nonlinear Science, University of California - San Diego, La
Jolla, CA 92093 - 0402

\bigskip

Nonlinear dynamical modeling of interaction between automatic and conscious
processes in the human brain is considered in terms of the quaternion
fields. The interaction is due, particularly, to the nonlinear firing rate
of neurons. Possible connection of consciousness with the general field
theory is indicated. A new type of symmetry between dynamics of real and
imaginary fields is pointed out.

\bigskip

From the physical-mathematical point of view consciousness ($C$) is a
mystery. It is desired to catch $C$ into some sort of equations. The common
knowledge is that $C$ is somehow connected with the electrochemical activity
in the brain. So, it seems logical to start with equations for these
processes. The brain activity revealed the regime of scale-similarity [1-3],
which is typical for systems with strong interaction of many degrees of
freedom (particularly, for turbulence [4]). Corresponding equations can be
formally written and are rather complicated. However, at this stage of our
understanding of $C$, the precise form of equations is not critical. One can
use various simplified models. The important question is how to connect
these equations with $C$? The $C$-processes are subjective and, as far as we
know, they can not be measured directly by the objective methods, which are
used for measuring the electrochemical (automatic) processes. At the same
time, there are reasons to believe that $C$-processes can interact with
automatic ($A$) processes. We need equations for $A$-fields and $C$-fields,
which interact despite the fact that $C$-field have a different nature and
can not be measured directly by the same methods as $A$-fields.

In recent papers [5-8] an approach to nonlinear dynamical modeling of
interaction between these two processes was presented. The idea is to use
the quaternion field with real and imaginary components representing $A$-
and $C$-processes. The subjective $C$ - experiences were divided into three
major groups: sensations ($S$), emotions ($E$) and reflections ($R$). Note,
that subjective $S$ should be distinguished from the automatic sensory input
into the neuron system of the brain [9]. The $A-C$ interaction is due to the
nonlinearity of the system. This approach was illustrated on the nonlinear
equation for the current density in the cortex.. The nonlinearity is
determined by the sigmoidal firing rate of neurons. Perspective for the
laboratory testing of this approach were also indicated as well as some more
general approaches [5-8].

To be specific, consider quaternion:

\begin{equation}
q=\alpha +i_{p}\psi _{p}  \tag{1}
\end{equation}%
Here $\alpha (t)$ is the average (spatially uniform) current density
perpendicular to the cortical surface, $\psi _{p}(t)$ represent the
indicated above ($S,E,R$) - effects and summation is assumed on repeated
subscripts from 1 to 3. The imaginary units $i_{p}$ satisfy condition:

\begin{equation}
i_{p}i_{s}=\varepsilon _{psr}i_{r}-\delta _{ps}  \tag{2}
\end{equation}%
where $\ \varepsilon _{psr}$ is the unit antisymmetric tensor and $\ \delta
_{ps}$ is the unit tensor. Formula (2) is a compact form of conditions: $%
i_{1}^{2}=i_{2}^{2}=i_{3}^{2}=-1,$ $i_{1}i_{2}=-i_{2}i_{1}=i_{3},$ $%
i_{2}i_{3}=-i_{3}i_{2}=i_{1},$ $i_{3}i_{1}=-i_{1}i_{3}=i_{2}.$

The model equation for the quaternion $q$ has the form [5-8]:

\begin{equation}
\frac{\partial q}{\partial t}+kq=f(q+\sigma )+\phi ,\text{ \ }\sigma
=s+i_{p}\varphi _{p}  \tag{3}
\end{equation}%
Here $k$ is the relaxation coefficient, $f$ represents the sigmoidal firing
rate of neurons [for example, $f(\alpha )=\tanh (\alpha )$], $\phi $
represents the external electromagnetic ($EM$) excitations. The quaternion \ 
$\sigma $ is the averaged sensory input, which has real component $s$ and
imaginary components $\varphi _{p}$ (for so-called extra-sensory effects, if
they exist).

For the case of spatially nonuniform $q(t,\mathbf{x}),$ $\sigma (t,\mathbf{x}%
)$ and $\phi (t,\mathbf{x})$ we can use more general equation, which include
typical propagation velocity of signals in the neuron system of the cortex $%
v $. Time differentiation of (3), simple algebra and addition a term with
the two-dimensional spatial Laplacian $\Delta $ gives [5-8]:

\begin{equation}
\frac{\partial ^{2}q}{\partial t^{2}}+(k+m)\frac{\partial q}{\partial t}%
+(km-v^{2}\Delta )q=(m+\frac{\partial }{\partial t})f(q+\sigma )+\frac{%
\partial \phi }{\partial t}  \tag{4}
\end{equation}%
where $m$ is an arbitrary parameter (see below). Real and imaginary
projections of (4) give a system of four partial differential equations for $%
\alpha $ and $\psi _{p}$. If we put $\psi _{p}=0$ and $\phi =0$, than
equation for $\alpha $ will be similar in spirit to equation used for
interpretation of EEG an MEG spatial patterns (see recent paper [10 ] and
references therein). In this context we have parameters: $k\sim m\sim v/l$,
where $l$ is the connectivity scale.

Returning, for simplicity, to (3) and using $f(\alpha )=\tanh (\alpha )$, we
have explicit projections [7]:

\begin{equation}
\frac{\partial \alpha }{\partial t}+k\alpha =\frac{\sinh [2(\alpha +s)]}{%
\cosh [2(\alpha +s)]+\cos (2\theta )}+\phi ,\text{ \ }\theta ^{2}=\theta
_{p}^{2},\text{ \ }\theta _{p}=\psi _{p}+\varphi _{p}  \tag{5}
\end{equation}

\begin{equation}
\frac{\partial \psi _{p}}{\partial t}+k\psi _{p}=\frac{\theta _{p}\theta
^{-1}\sin (2\theta )}{\cosh [2(\alpha +s)]+\cos (2\theta )},\text{ \ }p=1,2,3
\tag{6}
\end{equation}

Some general conclusions can be made without solving these equations.
Firstly, if $\psi _{p}(0)=0$ and $\varphi _{p}(t)\equiv 0$ (no extra-sensory
effects), than $\psi _{p}(t)\equiv 0$. In other words, consciousness,
according to these equations, can not appear from nothing. Secondly,
presence of $\psi _{p}\neq 0$ through field $\theta $ (if not nullified by $%
\varphi _{p}$) changes the measurable field $\alpha $, so the model is
testable. Thirdly, if the neuron system ($NS$) is not working ($f(\alpha
)\equiv 0$), then $\psi _{p}(t)$ will decay exponentially ($k>0$). Finally,
if $\psi _{p}(0)\neq 0$ and $NS$ is working, than evolution $\psi _{p}(t)$
depends on $\alpha (0)$ and is influenced externally by the sensory input
and by $EM$ excitations. The nonlinearity of the system suggests that the
efficiency of external influence depends not only on amplitude (of, say, $%
s(t)$ ), but also of the shape (spectral content).

Generally, $\psi _{p}(t,\mathbf{x})$ signifies the presence of imaginary
components of $EM$ field in the brain. Let us assume, as an adventure, that
such field can exist (at least, for a short time) in empty space (perhaps in
a brane [11], which encloses our ordinary 4D spacetime). Than, it seems
natural, that such field can propagate. According to the described model,
real and imaginary components of $EM$ field interact in the presence of
healthy $NS$. This interaction, apparently, is symbiotic and stable.
However, in special circumstances, when $NS$ is strongly disturbed, we can
expect shedding of a part of imaginary $EM$ field from this $NS$,
propagating and attracting to another $NS$ ( $\varphi _{p}$ - effect in the
model). In this way we can interpret observed instances of telepathy,
reincarnation ("past life memory" connected with fatal accidents) and
related effects, which can not be explained by traditional approaches. New
type of protocols have to be developed for laboratory testing of the $C$%
-modeling [5].

Interaction between $NS$`s by means of the field $\varphi _{p}(t,\mathbf{x})$
requires more general modeling. It will be very interesting to incorporate
fields $\psi _{p}(t,\mathbf{x})$ and $\varphi _{p}(t,\mathbf{x})$ into the
framework of contemporary field theory. Possible candidates are so-called
"ghosts" [11,12] with negative norm (energy). The negative norm corresponds
to imaginary field ($IF$). The string theorists are trying to get rid of
ghosts (with exception of Faddeev-Popov variety). But cleansing procedures
look rather artificial - for open bosonic string it requires to rise
spacetime dimension to $D=26$. Perhaps we should embrace $IF$, but in
different context. Could it be that brain tissue with healthy $NS$ attracts $%
IF$ and such attraction is used by Nature to create conscious beings? That
is a thrilling (trillion dollars) question.

It is not that we need the full-blown strings theory to deal with
consciousness. But, it is important to know that ($\psi ,\varphi $) -
fields, which we introduced phenomenologically, can be derived from the
first principles [13]. On another hand, recognition of $IF$ as legitimate
fields can help in development of the general field theory. Note, that
fields, which can be responsible for the accelerated expansion of the
Universe [12] (particularly, distributed sources [14]), potentially may also
produce $C$-effects. This is a little scary project. It leads, for example,
to a new type of symmetry - between real fields ($RF$) and $IF$. From the
"point of view" of $IF$-world our $RF$ are imaginary [15]. We will not
discuss here philosophical aspects of such symmetry.

From practical (medical, technological) point of view, the most important is
to know - what is so special about the brain tissue with healthy $NS$, which
attracts $IF$ and give them a life, so to speak? And related question - can
we artificially create a composite material, which will attract and support $%
IF$? The described above simple model can be a hint, a beginning of a long
journey in answering these questions. Numerical experiments with this model
are in progress and results will be compared with observations and reported
elsewhere.

\bigskip

\textbf{References}

\bigskip

[1] E. Novikov, A. Novikov, D. Shannahof-Khalsa, B. Schwartz, and J. Wright,
Scale-similar activity in the brain, Phys. Rev. E \textbf{56}(3), R2387
(1997)

[2] E. Novikov, A. Novikov, D. Shannahof-Khalsa, B. Schwartz, and J. Wright,
Similarity regime in the brain activity, Appl. Nonl. Dyn. \& Stoch. Systems
(Ed. J. Kadtke \& A. Bulsara), p. 299, Amer. Inst. Phys., N. Y., 1997

[3] E. S. Freeman, L. J. Rogers, M. D. Holms, D. L. Silbergelt, Spatial
spectral analysis of human electrocorticograms including alpha and gamma
bands, J. Neurosci. Meth. \textbf{95}, 111 (2000)

[4] E. A. Novikov, Infinitely divisible distributions in turbulence, Phys.
Rev. E \textbf{50}(5), R3303 (1994)

[5] E. A. Novikov, Towards modeling of consciousness, arXiv:nlin.PS/0309043

[6] E. A. Novikov, Quaternion dynamics of the brain, arXiv:nlin.PS/0311047

[7] E. A. Novikov, Manipulating consciousness, arXiv:nlin.PS/0403054

[8] E. A. Novikov, Modeling of consciousness, Chaos, Solitons \& Fractals,
in press (2005).

[9] A. R. Damasio, \emph{The Feeling of What Happens}, Haircourt Brace \&
Company, 1999

[10] V. K. Jirsa, K. J. Jantzen, A. Fuchs, and J. A. Kelso, Spatiotemporal
forward solution of the EEG and MEG using network modeling, IEEE Trans. Med.
Imaging, \textbf{21}(5), 497 (2002)

[11] C. V. Johnson, \emph{D - Branes}, Cambridge Univ. Press, 2003

[12] Y. Fujii and K. Maeda, \emph{The Scalar - Tensor Theory of Gravitation}%
, Cambridge Univ. Press, 2003

[13] Recent review of attempts to connect $C$ with quantum theory (without $%
IF$) can be found in L. P. Rosa and J. Faber, Quantum models of the mind:
Are they compatible with environment decoherence?, Phys. Rev. E \textbf{70},
031902 (2004)

[14] E. A. Novikov, Dynamics of distributed sources, Phys. of Fluids, 
\textbf{15} (9), L65 (2003)

[15] For example, multiplication of equation (3) by imaginary unit $i_{p}$
gives equation for a quaternion with real component $-\psi _{p}$ and
corresponding imaginary component $\alpha $.

\bigskip

\end{document}